\documentclass[twocolumn,english,aps,prb,showpacs]{revtex4}
\usepackage[T1]{fontenc}
\usepackage{amsmath,graphicx,amssymb,epsfig,babel,dsfont}
\usepackage{mathrsfs}
\usepackage{color}
\begin{document}

%\title{Non-adiabatic vs adiabatic heat pumping in many-body systems}
\title{Heat transfer mediated by the Berry-phase in non-reciprocal many-body systems}

\author{S.-A. Biehs}
\email{s.age.biehs@uni-oldenburg.de}
\affiliation{ Institut f\"{u}r Physik, Carl von Ossietzky Universit\"{a}t, D-26111 Oldenburg, Germany}
\author{P. Ben-Abdallah}
\email{pba@institutoptique.fr}
\affiliation{Laboratoire Charles Fabry, UMR 8501, Institut d'Optique, CNRS, Universit\'{e} Paris-Saclay, 2 Avenue Augustin Fresnel, 91127 Palaiseau Cedex, France.}

%

%\date{\today}

%\pacs{44.40.+a, 78.20.N-, 03.50.De, 66.70.-f}
\begin{abstract}
We investigate the adiabatic evolution of thermal state in non-reciprocal many-body systems coupled to their environment and subject to periodic drivings. In such systems we show that besides the dynamical phase a geometrical phase can exist and it drives the relaxation dynamic of the system. On the contrary to the dynamical phase which always pushes the system toward its equilibrium state we show that the geometric phase can speed up or reduce the speed of relaxation process. These results could have applications in the field of thermal management of complex systems. 
\end{abstract}

\maketitle

\section{Introduction}

Understanding and controlling the time evolution of the thermal state of a system in non-equilibrum situation is of tremendeous 
importance in physics. Many strategies have been implemented to date to actively control this evolution using an external driving. 
Hence, by modulating some intensive quantities, such as the temperature or the chemical potential, an additional flux to the primary 
flux induced by a temperature bias can be generated and used to control heat exchanges. This shuttling effect~\cite{Li_2008,Li_2009,Latella,Messina}
results from the variation of the local curvature of flux with respect to these parameters. When the system displays a negative differential 
thermal resistance (i.e.\ a negative curvature of flux), this effect can contributes to inhibit the primary flux and even can pump heat 
from the cold to the hot part of the system. A slow cycling modulation of control parameters near-topological singularities~\cite{Li_2021,Xu} 
such as exceptional points can also be used to enhance or inhibit energy exchanges within a system. Finally, the  spatio-temporal modulation 
of thermal properties, such as the thermal conductivity or the specific heat, in systems can give rise~\cite{Torrent} to an effective convective 
flux which superimposes to the diffusive flux. This leads to an apparent change of heat transport regime which can  be exploited to control  
heat flows in solids networks at mesoscopic and macroscopic scales. 
Beside these developpements the concept of geometric phase theorized by 
Berry~\cite{Berry} has been exploited to develop novel pumping strategies in quantum and classical systems. Inspired by the Thouless charge 
pumping~\cite{Thouless} heat pumping in solid-state systems has been proposed~\cite{Ren} to control heat flux in numerous classical and 
quantum systems~\cite{Ren2,Ren3}. 

In this work, we introduce a general theory to describe the temporal evolution of thermal state of arbitrary non-reciprocal many-body systems~\cite{Hanggi,Baowen_rmp,Biehs2021} close to their equilibrium state under the action of external periodic drivings. In the adiabatic limit we show that this modulation can be used to slightly modify the relaxation dynamics. We show that in this limit a geometrical phase can exist which adds to the dynamical phase. We discuss the necessary conditions for the existence of the geometrical phase and provide a general example in a two-body system interacting with an external bath as well an example for near-field heat transfer in a many-body system.

\section{Time evolution of thermal state in $N$-body systems}

To start let us consider a generic many-body system made with $N$  bodies in mutual interaction and in interaction with an external bath at temperature $T_b$. The time evolution of thermal state $\bold{T}=(T_1,...,T_N)$ of this system  under temporal driving is governed by an energy balance (master) equation of the general form
\begin{equation}
	C_i\frac{d{T}_i}{dt}=\underset{j\neq i}{\sum}\mathscr{P}_{j\rightarrow i}(\bold{T};T_b,t), \:\:i={1,...,N}.
	\label{Eq:dynamic}
\end{equation}
Here $\mathscr{P}_{j\rightarrow i}$ denotes the power received by the $i^{th}$ element from the $j^{th}$ element within the system and $C_i$  its heat capacity. Close to the equilibrium state $\bold{T}_{eq}=(T_b,...,T_b)^t$ the net power can be linearized and expressed in term of pairwise thermal conductance 
\begin{equation}
  G_{ij}=\underset{T_j \rightarrow T_i}{\rm lim} \frac{\mathscr{P}_{j\rightarrow i}}{T_j-T_i}
\end{equation}
and of conductance 
\begin{equation}
  G_{ib}=\underset{T_b \rightarrow T_i}{\rm lim} \frac{\mathscr{P}_{b\rightarrow i}}{T_b-T_i}
\end{equation}
between the bath and each element. In this approximation equations (\ref{Eq:dynamic}) can be recast in the matrix form
\begin{equation}
	\mathds{C}(t)\frac{d\bold{T}}{dt}=\hat{\mathds{G}}(t)\bold{T}+\bold{S}(t),
	\label{Eq:dynamic3}
\end{equation}
where $\hat{\mathds{G}}(t)$ denotes the conductances matrix with 
\begin{equation}
  \hat{G}_{ij}=-\underset{j\neq i}{\sum}(G_{ij}+G_{ib})\delta_{ik}-(1-\delta_{ik})G_{ik},  
\end{equation}
$\mathds{C}={\rm diag}(C_1,...,C_N)$ is the heat capacity matrix and $\bold{S}(t)$ is the source vector induced by the bath (i.e. $S_i=G_{ib}T_b$). The time evolution of thermal state is given by the following expression
\begin{equation}
  \bold{T}(t)=\mathds{U}(t_0,t)\bold{T}_0+\int_{t_0}^{t}\mathds{U}(\tau,t)\mathds{C}^{-1}(\tau) \bold{S}(\tau)d\tau.
\label{Nbody_solution}
\end{equation}
Here $\mathds{U}$ denotes the propagator of differential system (\ref{Eq:dynamic3}) and $\bold{T}_0=(T_1(t_0),...,T_N(t_0))^t$ is the initial thermal state. In reciprocal systems (i.e. $\hat{G}_{ij}=\hat{G}_{ji}$) the fundamental matrix can be expressed in term of the exponential of the conductance matrix so that the thermal state (\ref{Nbody_solution}) can easily be calculated. On the other hand, in non-reciprocal systems (i.e. $\hat{G}_{ij}\neq\hat{G}_{ji}$), the situation is more tricky and no simple expression of thermal state can be derived. Nevertheless, following Garrisson and Wright~\cite{Garrison} we can, in the adiabatic appoximation, seek a solution of this system by expanding it over the basis $\{\boldsymbol{\varphi}_{\perp}\}_n$ of eigenstates of $\hat{\mathds{G}}$. To proceed we first recast the master equation as
\begin{equation}
	\frac{d\bold{\tilde{T}}}{dt}=\tilde{\mathds{G}}(t)\bold{\tilde{T}},
	\label{Eq:dynamic4}
\end{equation}
by setting $\bold{\tilde{T}}=(\bold{T},T_b)^t$ and
\begin{equation}
 \tilde{\mathds{G}}=\left(\begin{array}{ccc}
\hat{\mathds{G}} & \vdots & \left(\begin{array}{c}
G_{1b}\\
\vdots\\
G_{Nb}
\end{array}\right)\\
\cdots & \cdots & \cdots\\
\left(\begin{array}{ccc}
0 & \cdots & 0\end{array}\right)& \vdots & 0
\end{array}\right).
\label{block}
\end{equation}
Note, that for convenience $\hat{\mathds{G}}$ is now the conductance matrix normalized by the heat capacities.
Then we seek a solution of Eq.(\ref{Eq:dynamic4}) using the following expansion
\begin{equation}
	\bold{\tilde{T}}(t)=\overset{N+1}{\underset{i=1}{\sum}}c_i(t)e^{\gamma_{di}(t)}\boldsymbol{\varphi}_i(t),
	\label{solution}
\end{equation}
where $\gamma_{di}(t)=\int_{0}^{t}\lambda_i(\tau)d\tau$ is the dynamical phase associated the eigenvalue $\lambda_i$ of $\tilde{\mathds{G}}$ while $\boldsymbol{\varphi}_i$ is its eigenvector. It is easy to check that $\lambda_{N+1}=0$ is an eigenvalue  with the corresponding eigenvector $\boldsymbol{\varphi}_{N+1}=\bold{1}_{N+1}=(1,...,1)^t$. It turns out that the temperatures vector takes the form
\begin{equation}
	\bold{\tilde{T}}(t)=\overset{N}{\underset{i=1}{\sum}}c_i(t)e^{\gamma_{di}(t)}\boldsymbol{\varphi}_i(t)+c_{N+1}\bold{1}_{N+1}.
	\label{solution1}
\end{equation}

Since $\bold{\tilde{T}}\rightarrow (\bold{T}_{eq},T_b)^t$ after a sufficiently long time ($\hat{\mathds{G}}$ being strictly definite negative) we can write the solution of Eq.(\ref{Eq:dynamic4}) as
\begin{equation}
	\bold{\tilde{T}}(t)=\overset{N}{\underset{i=1}{\sum}}c_i(t)e^{\gamma_{di}(t)}\boldsymbol{\varphi}_i(t)+T_b\bold{1}_{N+1},
	\label{solution2}
\end{equation}
Taking the projection of this temperature vector on the subspace $\mathbf{e}_1\otimes...\otimes \mathbf{e}_N$ generated by the basis vectors $\mathbf{e}_i$ of the canonic base, we get finally the thermal state
\begin{equation}
	\bold{T}(t)=\overset{N}{\underset{i=1}{\sum}}c_i(t)e^{\gamma_{di}(t)}\boldsymbol{\varphi}_{\perp i}(t)+T_b\bold{1}_{N},
	\label{solution3}
\end{equation}
where according to the form of the block matrix defined in (\ref{block}) $\boldsymbol{\varphi}_{\perp i}$ is the $i^{th}$ eigenvector of $\hat{\mathds{G}}$.

\section{Adiabatic limit and geometrical phase}

Inserting the solution in Eq.~(\ref{solution3}) into the master equation (\ref{Eq:dynamic3}) we obtain, after removing the subscript $\perp$ for readability reasons, the relations
\begin{equation}
	\overset{N}{\underset{i=1}{\sum}}\partial_t(c_i e^{\gamma_{di}}\boldsymbol{\varphi}_ i)=\overset{N}{\underset{i=1}{\sum}}c_i(t)e^{\gamma_{di}}\hat{\mathds{G}}\boldsymbol{\varphi}_ i+T_b\hat{\mathds{G}}\bold{1}_N + \bold{S}.
	\label{Berry1}
\end{equation}
As $T_b\tilde{\mathds{G}}\bold{1}_N=-\bold{S}$ and $\hat{\mathds{G}}\boldsymbol{\varphi}_ i=\lambda_i \boldsymbol{\varphi}_i$ we have equivalently
\begin{equation}
	\overset{N}{\underset{i=1}{\sum}}(\dot{c}_i e^{\gamma_{di}}\boldsymbol{\varphi}_ i+c_i e^{\gamma_{di}}\dot{\boldsymbol{\varphi}}_ i)=0.
	\label{Berry2}
\end{equation}
Multiplying this relation by $\boldsymbol{\psi}_j$ the $j^{th}$ eigenvector of the transpose matrix $\hat{\mathds{G}}^t$ and using the biorthogonality relations $\boldsymbol{\psi}_i \cdot \boldsymbol{\varphi}_j = \delta_{ij}$ between the eigenvectors of $\hat{\mathds{G}}$ and $\hat{\mathds{G}}^t$ we find 
\begin{equation}
	\dot{c}_j e^{\gamma_{dj}}+\overset{N}{\underset{i=1}{\sum}}c_i e^{\gamma_{di}} \boldsymbol{\psi}_j\cdot \dot{\boldsymbol{\varphi}}_i = 0.
	\label{Berry3}
\end{equation}
Now, assuming a slow (adiabatic) variation of external control parameters with respect to time this system simplifies into
\begin{equation}
	\dot{c}_j+c_j \boldsymbol{\psi}_j \cdot \dot{\boldsymbol{\varphi}}_j = 0,
	\label{Berry4}
\end{equation}
whose the solution reads
\begin{equation}
	c_j(t)=\alpha_j e^{\gamma_{gj}(t)}, 
	\label{c_i}
\end{equation}
with 
\begin{equation}
  \gamma_{gj}(t)=-\int_0^t\!\! {\rm d}\tau \,\boldsymbol{\psi}_j \cdot \dot{\boldsymbol{\varphi}}_j (\tau)
\end{equation}
the so called geometric phase. Hence, in the adiabatic limit the thermal state can be written as
\begin{equation}
	\bold{T}(t)=\overset{N}{\underset{i=1}{\sum}}\alpha_i e^{\gamma_{di}(t)}e^{\gamma_{gi}(t)}\boldsymbol{\varphi}_i(t)+T_b\bold{1}_{N}.
	\label{solution4}
\end{equation}
This expression is the classical analog of the Berry's result to arbitrary non-reciprocal many body systems. 
The constants $\alpha_i$ can be readily calculated from the initial thermal state $\bold{T}_0$.

\begin{figure}%[!hbt]
%\centering
\includegraphics[angle=0,scale=0.4]{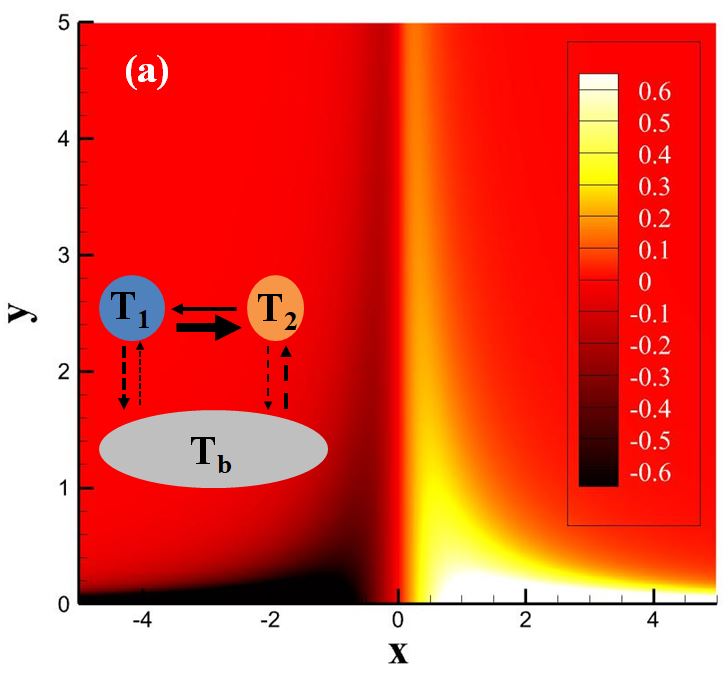}
\includegraphics[angle=0,scale=0.4]{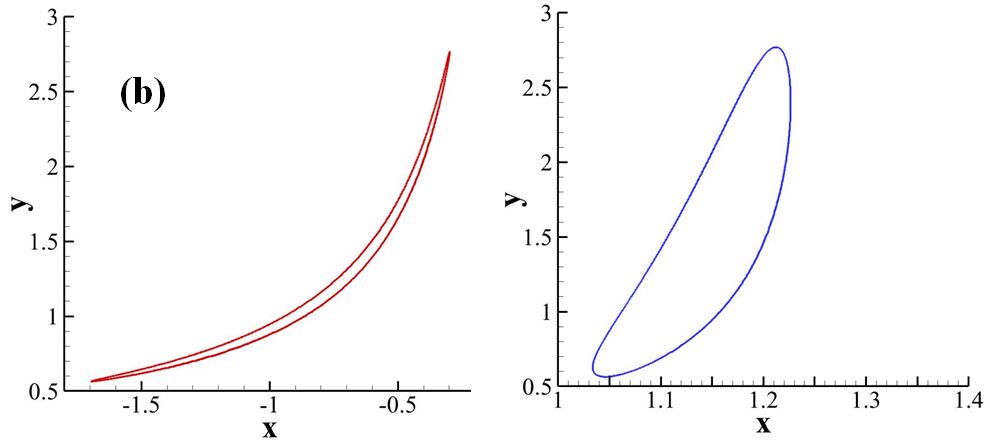}
\includegraphics[angle=0,scale=0.4]{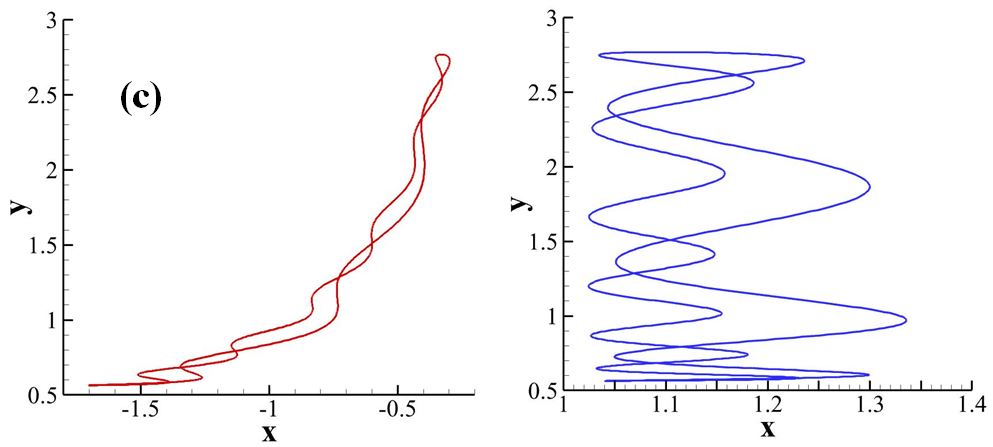}
\includegraphics[angle=0,scale=0.42]{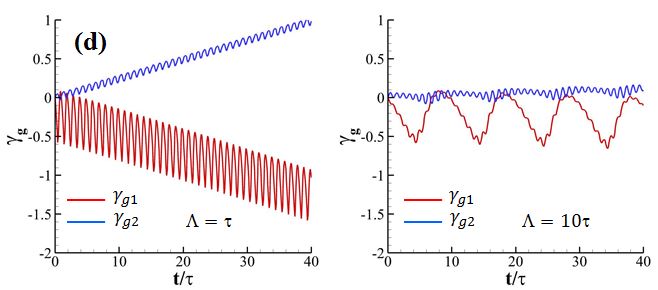}
\caption{(a) Non-vanishing component of the effective magnetic field in the parameter space $(x,y)$. The inset is a sketch of studied system made with two solids in mutual interaction with asymmetric conductances and in interaction with an external bath. (b) Trajectory in the parameter space associated to two different eigenvalues of $\hat{\mathds{G}}$ when $\Lambda=\tau$ and (c) $\Lambda=10\tau$. (d) Cumulative geometric phases with $\tau=1\:s$.}
\label{Fig_1}
\end{figure} 

\section{A simple toy model}

To show to what extent the flux is affected by the geometrical phase we consider below a generic system as sketched in (Fig.1-a) made of two objets in mutual interaction and in interaction with the external bath. In this case the conductance matrix takes the form
\begin{equation}
  \hat{\mathds{G}}=\left(\begin{array}{cc}
   -a & G_{12} \\
   G_{21} & -b\\
  \end{array}\right),
  \label{2body}
\end{equation}
with $a=G_{12}+G_{1b}$ and $b=G_{21}+G_{2b}$. The eigenvectors of $\hat{\mathds{G}}$ and of its transpose are 
\begin{equation}
	\boldsymbol{\varphi}_i= \begin{pmatrix} 1 \\ \frac{G_{21}}{b+\lambda_i} \end{pmatrix} \quad\text{and}\quad \boldsymbol{\psi}_i =  \frac{1}{\beta_i} \begin{pmatrix} 1 \\ \frac{G_{12}}{b+\lambda_i} \end{pmatrix}, 
\end{equation}
where 
\begin{equation}
  \lambda_i=\frac{1}{2}{\rm tr}\hat{\mathds{G}}\pm \biggl[\frac{1}{4}{\rm tr}^2\hat{\mathds{G}}-{\rm det} \hat{\mathds{G}}\biggr]^{1/2}
\end{equation}
denotes the (real) eigenvalues of $\hat{\mathds{G}}$, $\beta_i$ being calculated from the normalization relation $\boldsymbol{\varphi}_i \cdot \boldsymbol{\psi}_i = 1$. By setting $x_i=\frac{G_{21}}{b+\lambda_i}$ and $y_i=\frac{G_{12}}{G_{21}}$ it follows that the geometric phase accumulated between the initial instant up to time $t$ is 
\begin{equation}
	\gamma_{gi}(t)=-\int_{t_0}^t \!\!{\rm d} \tau \,\frac{x_i \dot{x_i}y_i}{1+x_i^2 y_i}.
  \label{berry_phase}
\end{equation}

Now, suppose that $\hat{\mathds{G}}(t)=\hat{\mathds{G}}(\bold{R}(t))$, where $\bold{R}=(R_1,R_2)^t$ denotes a set of control parameters and that there is a time $T$ for which $\bold{R}(t_0+T)=\bold{R}(t_0)$. Hence, we consider a closed trajectory in parameter space. Then the accumulated geometrical phase in Eq.~(\ref{berry_phase}) during this interval $T$ can be written as
\begin{equation}
	\gamma_{gi}(T)=\oint_{\partial S} {\rm d}\bold{R} \cdot \bold{A}_i,
  \label{berry_phase2}
\end{equation}
where 
\begin{equation}
	\bold{A}_i = (\nabla_{\bold{R}} \boldsymbol{\varphi}_i)^t  \cdot \boldsymbol{\psi}_i
\end{equation}
is analog to a vector potential whose associated magnetic field is $\bold{B}_i=\nabla\times \bold{A}_i$. Here $\partial S$ is the path of the trajectory in parameter space. Using the Stoke theorem with the oriented surface $S$ enclosed by the countour $\partial S$ generated by the change of parameter $\bold{R}$ during the interval $[t_0,t_0+T]$ we can express the geometric phase in term of $\bold{B}_i$ as
\begin{equation}
	\gamma_{gi}(T)=\int_S {\rm d}\bold{S} \cdot\bold{B}_i
  \label{berry_phase3}
\end{equation}
where ${\rm d}\mathbf{S}$ is the infinitesimal oriented surface element on $S$.

These expressions allow us to determine the vector potential and magnetic field in our two-body system for any trajectory $\mathbf{R}$. With Eq.~(\ref{berry_phase}) for one period $T$ we find 
\begin{equation}
  \mathbf{A}_i = \frac{x_i y_i}{\beta_i} \bigl( \nabla_{\mathbf{R}} x_i \bigr).
\end{equation}
Note that this expression is not gauge invariant as it is typical for vector potentials. For example, also the expression 
\begin{equation}
  \mathbf{A}_i' = - \frac{x_i^2}{2 \beta_i} \bigl( \nabla_{\mathbf{R}} y_i \bigr)
\end{equation}
is a valid representation of the vector potential. On the other hand, the magnetic field is gauge invariant, and consequently we obtain for both $\mathbf{A}_i$ and $\mathbf{A}_i'$ the magnetic field
\begin{equation}
  \mathbf{B} = - (\nabla_{\mathbf{R}} x_i) \times  \bigl( \nabla_{\mathbf{R}} y_i \bigr) \frac{x_i}{\beta_i^2}.
\end{equation}
Interestingly, from this expression it is obvious that $\mathbf{B} = \mathbf{0}$ if $y_i = 1$, i.e.\ if $G_{12} = G_{21}$. Hence, non-reciprocity is a necessary condition for the appearance of a Berry phase. Furthermore, the magnetic field vanishes if $\nabla_{\mathbf{R}} x_i$ and $\nabla_{\mathbf{R}} y_i$ are parallel or antiparallel. This happens for a constant background conductance if $\nabla_{\mathbf{R}} G_{12}$ and $\nabla_{\mathbf{R}} G_{21}$ are parallel or antiparallel, i.e.\ if for example $G_{12}$ and $G_{21}$ change in phase with the external parameter change. Therefore, if $G_{12}$ and $G_{21}$ are in phase or phase-shifted by $\pi$, then there is no Berry phase.

%Taking $\bold{R}=(x_i,y_i)^t$ as the natural set of driving parameters we immediately get
%$\bold{A}_i=-\frac{1}{1+x_i^2 y_i} ^t(x_i y_i,0,0)$ and $\bold{B}_i=\frac{x_i}{(1+x_i^2 y_i)^2} ^t(0,0,1)$.

The above statements are very general. Now, we take a conrete path in the parameter space by choosing the two natural parameters $x_i$ and $y_i$ for a path embedded in $\mathds{R}^3$, i.e.\ $\mathbf{R}(t) = (x_i(t),y_i(t),0)^t$. Then the magnetic field is
\begin{equation}
  \mathbf{B} = - \frac{x_i}{\beta_i^2} \mathbf{e}_z.
\end{equation}
%This quantitiy is shown in Fig.~\ref{Fig:Bfield}. From this figure it can be concluded that any cyclic external parameter change resulting in a closed path of $\mathbf{R}(t) = (x_i(t),y_i(t),0)^t$ enclosing a non-vanishing area $S$ in either the first or the fourth quadrant of the x-y plane will therefore result in a non-vanishing Berry phase.
In Fig.~1 we plot the non-vanishing component of $\bold{B}_i$ in the space of control parameters as well as the trajectories defined  by the parametric curve $\mathbf{R} = (x_i(t),y_i(t))^t$. It is clear from expresion (\ref{berry_phase3}) that the geometric phases $\gamma_{gi}(T)$ depends intimely on the the shape of closed contour generated by these trajectories during one oscillation period. 

In Fig. 1-b and 1-c we show some typical trajectories followed by these parameters during one cycle of modulation and the  corresponding cumulative geometric phase (Fig.1-d) when the exchanges conductances obey to the following variation laws
\begin{align}
	G_{12}&=G+\delta G\:\cos\biggl(\frac{2\pi}{\Lambda} t\biggr), \label{conductances1}\\
	G_{21}&=H+\delta H\: \cos\biggl(\frac{2\pi}{\Lambda} t+\theta\biggr),\\
	G_{1b}&=g+\delta g\: \cos\biggl(\frac{2\pi}{\tau} t\biggr),\\
	G_{2b}&=h+\delta h\: \cos\biggl(\frac{2\pi}{\tau} t+\theta\biggr).
\label{conductances4}
\end{align}
\begin{figure}%[!hbt]
%\centering
\includegraphics[angle=0,scale=0.5]{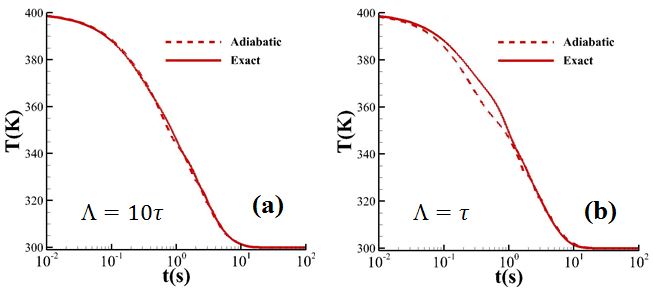}
\includegraphics[angle=0,scale=0.5]{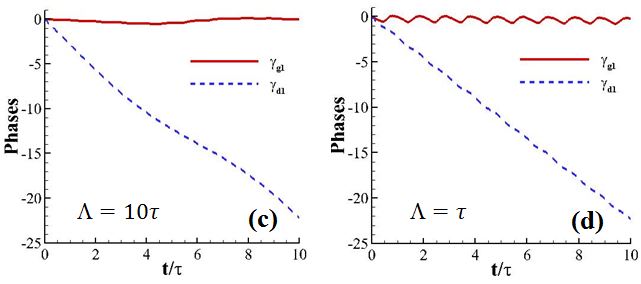}
	\caption{Approximate (adiabatic) and exact relaxation dynamics of hot bodies in a two body system driven by the conductance matrix (\ref{conductances1})-(\ref{conductances4}) with $G=1$, $H=0.8$, $g=0.5$, $h=0.3$, $\delta G=0.5$, $\delta H=0.4$, $\delta g=\delta h=0.1$ and $\theta=\pi/2$. The initial temperature is $T_1(0)=400\:K$   while the external bath is set at $T_b=300\:K$ in the case (a) $\Lambda=10\tau$ and (b) $\Lambda=\tau$ with $\tau=1\:s$. (c)-(d) Cumulated geometric and dynamic phases.} 
\label{Fig_2}
\end{figure}
The numerical results show clearly the sensitivity of geometric effect to the multiperiodicity of driving. When $\Lambda\sim\tau$ and $\Lambda\rightarrow \infty$  the integration surface reduces to a single point so that the Berry phase vanishes. On the contrary, with a finite period $\Lambda$ the closed countour delimits a non-vanishing area and this countour is simple  (without crossing point)  provide that $\Lambda=\tau$ whereas in the case of multiperiodic driving (i.e. $\Lambda\neq\tau$) this countour presents, in general,  several crossing points and it can be decomposed into several loops which are traveled either in clockwise or anti-clockwise direction. In the first case if a loop is in the first quadrant (i.e. $x_i>0$ and $y_i>0$) and the countour is browsed in clockwise direction resulting in a positive contribution to the geometric phase $\gamma_i$. In other words, during this period the geometric phase tends to insulate the different parts of the system. On the other hand, if the loop is browsed in the opposite direction the generated geometric phase is negative and the relaxation process is accelerated. This phenomenon is analog to the coiled light mechanism discovered by Chiao et al. \cite{Chiao1,Chiao2}  where the parameters space corresponds to the light polarization state. 

\begin{figure}%[!hbt]
\includegraphics[angle=0,scale=0.6]{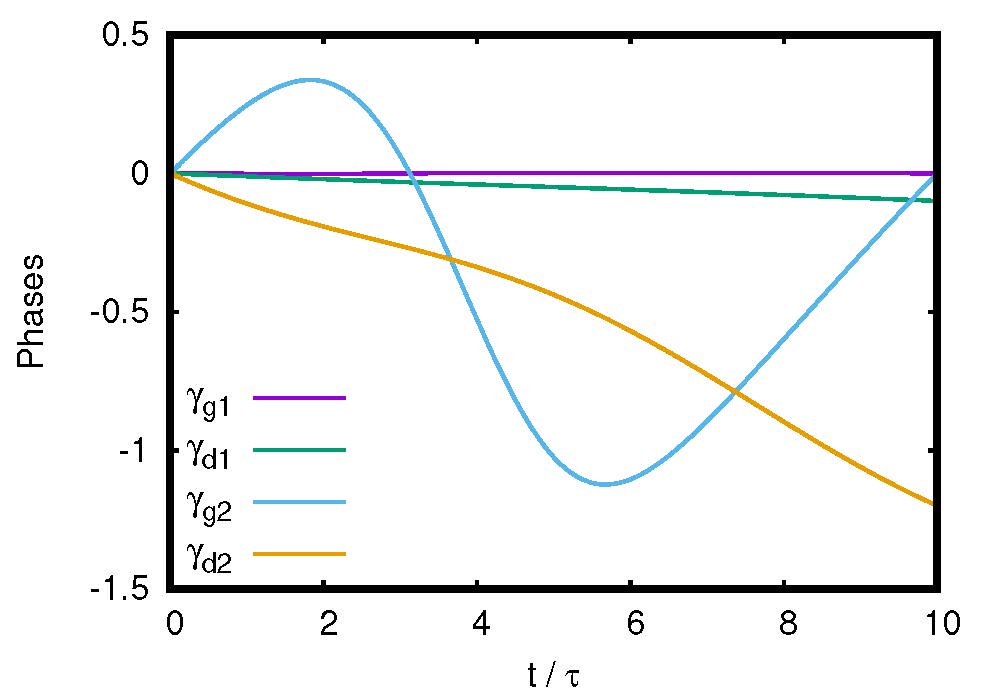}
	\caption{Geometrical and dynamical phases for the model as in Fig.~\ref{Fig_2}-c for $\Lambda = 10\tau$ but with $G=0.01$, $H=0.1$, $g=0.01$, $h=0.01$, $\delta G=0.005$, $\delta H=0.05$, $\delta g= 0.1g$, $\delta h=0.1h$, and $\theta=\pi/2$.}
\label{Fig_neu}
\end{figure}

Finally, the relaxation dynamic is plotted in Fig.~2 when $\Lambda=10\tau$ and $\Lambda=\tau$ both using the approximate adiabatic solution in Eq.~\ref{solution4} and a numerical exact solution of the energy balance equation in Eq.~(\ref{Eq:dynamic}) using a Runge-Kutta (RK) type method. When the oscillation period $\Lambda$ of main coupling channel is sufficiently large compared to the relaxation time of system we see (Fig.~2-a) that the relaxation dynamics is properly described by the adiabatic approximation~(\ref{solution3}). On the other hand, for shorter oscillations we observe in Fig.~2-b a deviation between this approximation and the exact evolution calculated with the RK method. Notice that, the cumulated dynamic phases (always negative) being, at large time scale,  larger than the cumulated geometric phase as shown in Fig.~2-c and Fig.~2-d it always pushes the system toward its equilibrium state. On the other hand, the geometrical phase, which can be either positive or negative, has an oscillatory character and its accumulation after one or several periods can be either positive or negative so that they can speed up or speed down the relaxation process. Unfurtunatly, this geometric effect cannot be persistent in time and the system finishes at the end to be driven only by the dynamical phase. In our toy model  the two Berry phases $\gamma_{gi}(\Lambda)$ become much smaller than the dynamical phases $\gamma_{di}(\Lambda)$ after one period $\Lambda=10\tau$ but during the first cycle they can be of the same order of magnitude and even much larger as visualized in Fig.~\ref{Fig_neu}. Optimization procedures will certainly be able to find the maximal Berry phases in this simple system and more generally in arbitrary many-body systems. However, this problem goes far beyond the scope of the present work.     

\section{conclusion}
In summary, although the physics of non-reciprocal systems remain today largely ellusive, the results introduced in this work highlight the peculiarities of relaxation processs for this systems when they are driven by periodic external actuations. On the contrary to reciprocal systems, the presence of a geometrical phase superimposes to the dynamical phase and has the potential  to significantly alter the relaxation dynamic of systems. We have shown that this phase can be used either to  accelerate or reduce the speed of relaxation. We hope that these preliminary  results will stimulate research on the thermal control of  non-reciprocal systems. On a theoretical point of view it would be interesting to explore the role played by the dissipation mechanisms induced by the external driving as well as the potential of multispectral drivings on the relaxation dynamics. The non-adiabatic control of these systems remains also a challenging problem. These problems will be addressed in subsequent studies. 

\begin{acknowledgments}
P.B.-A. acknowledges support from the Agence Nationale de la Recherche in France through the NBodheat project (ANR-21-CE30-0030-01). S.-A.\ B.\ acknowledges support from Heisenberg Programme of the Deutsche Forschungsgemeinschaft (DFG, German Research Foundation) under the project No.\ 404073166. This research was supported in part by the National Science Foundation under Grant No. NSF PHY-1748958.
\end{acknowledgments}

\end{document}